\documentstyle[12pt]{article}

\newcommand{\be}{\begin{equation}}
\newcommand{\ee}{\end{equation}}

\newcommand{\bea}{\begin{eqnarray}}
\newcommand{\eea}{\end{eqnarray}}

\newcommand{\p}{\partial}

\newcommand{\nn}{\nonumber \\}
\newcommand{\f}{\frac}
\newcommand{\w}{\wedge}
\newcommand{\ra}{\rightarrow}
\begin{document}

\thispagestyle{empty}

\begin{flushright}
\bf{arXiv:0808.3232}
\end{flushright}
\begin{center} \noindent \Large \bf
Towards Gravity solutions of AdS/CMT
\end{center}

\bigskip\bigskip\bigskip
\vskip 0.5cm
\begin{center}
{ \normalsize \bf  Shesansu Sekhar Pal}\\
\vskip 0.5cm

{ Barchana, Jajpur, 754081, Orissa, India \\
\vskip 0.5 cm

\sf shesansu${\frame{\shortstack{AT}}}$gmail.com}
\end{center}
\centerline{\bf \small Abstract} In this short note, we have
generalized and constructed  gravity solutions with two
``exponents'' {\it a la} Kachru, Liu and Mulligan. The coordinate
system that is used to construct the gravity solution is useful when
$b$ vanishes. It means we can describe the theory having only the
temporal scale invariance apart from the combination of both
temporal and spatial scale invariance. The two point correlation
function of the scalar field in the mass less limit is computed in a
special case that is $a/b=2$.

\newpage
Recently, it has been becoming very interesting to understand the gravity duals
to 2+1 dimensional CFT's so as to understand the strongly coupled behavior
of these CFT's, which may resemble some of the systems that we know in
condensed matter theories (CMT).
In this context several gravity solutions have been
generated \cite{klm}, \cite{son}-\cite{ssp}
and more solutions need to be constructed so as to understand
better the dual field theories at strong coupling.

The way to understand these systems is by constructing new gravity solutions
with specific symmetry group. The scaling symmetry that we shall
consider is
\be\label{scaling}
t \ra \lambda^a t,~~~(x,y)\ra \lambda^b (x,y),~~~ r \ra \f{r}{\lambda},
\ee

with the condition that we do not want to break the isotropicity
along $x$ and $y$ directions. The constraint on  the parameters $a$
and $b$ are $a \geq b$, and $a b \geq 0$. When $ab=0$, we shall only
take $b=0$, but not $a$. The metric that remains invariant under
this symmetry is \be \label{metric}
ds^2=L^2\bigg(-r^{2a}dt^2+r^{2b}(dx^2+dy^2)+\f{dr^2}{r^2} \bigg).
\ee For $a=b\neq 0$, the system has  SO(2,~3) symmetry group and is
described as $AdS_4$ space time. This particular choice of the metric
makes the time reversal manifest without the need to depend on the
parity of the other directions. Moreover, it is useful to deal with
for vanishing  $b$. Which means we can study temporal scale
invariance in this choice of coordinate system as opposed to the
coordinate system used in \cite{klm}.

For $a=2$ and $b=1$, the  field theory action, which preserves the
scaling symmetry,  may be written as \cite{hls} and \cite{gg}
\be\label{field_action} S_C=\f{1}{2}\int d^2x
dt[(\p_t\chi)^2-K(\nabla^2\chi)^2] \ee where $K$ is a constant and
describes a line of fixed points. If we break the rotational
symmetry in the $(x,y)$
 plane of eq(\ref{metric})
then the dual field theory action could be \cite{vbs}
\be
S=S_C+S_{int}
\ee
where
\bea
S_C&=&\f{1}{2}\int d^2x dt\bigg((\p_t\chi)^2-K[(\nabla^2\chi)^2+
4\sigma(\p^2_x\chi)(\p^2_y\chi)]\bigg),\nn
S_{int}&=&\int d^2x dt \bigg(\f{u}{4}[(\p_x\chi)^4+(\p_y\chi)^4]+\f{v}{2}(\p_x\chi)^2(\p_y\chi)^2\bigg)
\eea
and the isotropic point corresponds to $\sigma=0$ and $u=v$.

Now if we want to have a symmetry like eq(\ref{scaling}) for generic
$a$ and $b$ then the simplest dual field theory action, may be
described as \be\label{gen_scale_inv} S=\f{1}{2}\int d^2x
dt[(\p_t\chi)^{\alpha}-{\tilde K}(\nabla^2\chi)^{\beta}] \ee where
\be \alpha=\f{2b+a}{a}=1+\f{2b}{a},~~~\beta=\f{2b+a}{2b}=1+\f{a}{2b}
\ee

The action eq(\ref{gen_scale_inv}) has got the same structure
as in eq(\ref{field_action}) i.e. quadratic
in fields and the computation of the 2-pt correlation function becomes
Gaussian for a very specific choice of $a$ and $b$ that is $\f{a}{b}=2$.

Till now, the field theory action consistent with the scaling
symmetry eq(\ref{scaling}) that we have been discussing contains
first derivative in time and second derivative in space coordinates.
Now, the question arises: can we have an action that contains both
second order derivatives in time and space coordinates? The answer
to this is: yes\footnote{ $S=\f{1}{2}\int d^2 xdt[(t
\p^2_t\chi)^{\alpha}- {\tilde K}(\nabla^2\chi)^{\beta}]$, but we do
not know whether this kind of action is useful or not.}. \be
S=\f{1}{2}\int d^2 xdt[(\p^2_t\chi)^{\alpha}-{\hat
K}(\nabla^2\chi)^{\beta}] \ee where \be
\alpha=\f{2b+a}{2a}=\f{1}{2}+\f{b}{a},~~~\beta=\f{2b+a}{2b}=1+\f{a}{2b}
\ee and again for $a/b=2$, the action is not any more quadratic in
fields or for any other value of $a/b$. Hence, the most general
action\footnote{As far as time reversal is concerned, the Lagrangian
density is invariant but not the action, which changes sign by an
overall factor and under space parity the action remains invariant.}
consistent with the scaling symmetry and  quadratic in fields is
eq(\ref{gen_scale_inv}).

The metric eq(\ref{metric}) is non-singular and is well defined
everywhere except at the origin $r=0$, as it is not geodesically
complete, for $b\neq 0$ in \cite{klm} and possibly be the same even
for $b=0$. However, the coordinate invariant quantities that are
displayed in eq(\ref{invariant_quantity}) says that we can make
these quantities as small as we want by tuning the size i.e. $L$.
The action that generates such a solution can be obtained from the
action written in \cite{klm}, which is a system containing gravity
and fluxes of 2-form and 3-form type,  as the relevant degrees of
freedom

\be \label{action} S=\f{1}{2\kappa^2}\int d^4 x \sqrt{-g}
(R-2\Lambda)-\f{1}{4\kappa^2}\int (F_2\w\star F_2+F_3\w\star
F_3)+\f{c}{2\kappa^2}\int B_2\w F_2, \ee where $F_3=dB_2$ and  $c$ is
the topological coupling. The equations of motion  that follows from
it are

\bea &&d\star F_2=-c F_3,~~~d\star F_3=c F_2,\nn &&G_{MN}+\Lambda
g_{MN}=\f{1}{2}\sum_{p=2,3}\f{1}{p!} \bigg(p F_{MM_2M_3\cdots
M_p}F^{NM_2M_3\cdots M_p}-\f{1}{2}g_{MN}F_{M_1M_2\cdots
M_p}F^{M_1M_2\cdots M_p} \bigg)\nn \eea where $G_{MN}$ is the
Einstein tensor. Ansatz to the fluxes consistent with the scaling
symmetry are \cite{klm} \be F_2=\f{A L^2}{r^{1-a}}dr\w
dt,~~F_3=\f{BL^3}{r^{1-2b}}dr\w dx\w dy \ee where for the 2-form
it's only electric field and for 3-form it's magnetic type. It is
easy to check that these fluxes obey Bianchi identities:
$dF_2=0,~~dF_3=0$ even for A=A(r) and B=B(r).

Solving the equations of motion associated to these fluxes gives the
restriction \bea\label{flux_constraint} c B(r)&=&\f{r^{1-2b}}{L
}\f{d}{dr}(A(r)r^{2b}),\nn c A(r)&=&\f{r^{1-a}}{L} \f{d}{dr}(B(r)
r^a) \eea

From the  metric components as written in eq(\ref{metric}), we get
\bea G_{tt}+\Lambda g_{tt}&=&-r^{2a}(3b^2+L^2 \Lambda)\nn
G_{xx}+\Lambda g_{xx}&=&G_{yy}+\Lambda g_{yy}=r^{2b}(a^2+b^2+ab+L^2
\Lambda)\nn G_{rr}+\Lambda g_{rr}&=&\f{2ab+b^2+L^2\Lambda}{r^2} \eea

Finally, the equation of motion of the metric components gives
\bea\label{eom_metric} L^2(A^2+B^2)&=&-4 (3 b^2+L^2\Lambda)\nn
L^2(A^2+B^2)&=&4 (a^2+ b^2+ab+L^2\Lambda)\nn L^2(B^2-A^2)&=&4 (2ab+
b^2+L^2\Lambda), \eea where these equations arise from  tt, xx and
rr components, respectively. From these equations, we see that the
right hand side of it are constants, which means the functions A(r)
and B(r) better be constants. It means, eq(\ref{flux_constraint})
gives \be\label{flux_relation} \f{A}{B}=\f{a}{cL}=\f{cL}{2b} \ee

Now solving the equations eq(\ref{eom_metric}) gives
\bea\label{solutions} A^2&=&\f{2a(a-b)}{L^2},\nn
B^2&=&\f{4b(a-b)}{L^2},\nn \Lambda&=&-\f{a^2+ab+4b^2}{2L^2} \eea It
just follows that the coupling $c$ is not any more arbitrary but is
related to the exponents $a$ and $b$ as \be c^2 L^2=2 a b. \ee

Some of the interesting properties about the solution
eq({\ref{metric}}) \bea\label{invariant_quantity} R({\rm scalar
~~curvature})&=&-\f{2}{L^2}(a^2+2ab+3b^2)=-\kappa^2 T^M_M,\nn
R_{MN}R^{MN}&=&\f{2}{L^4}(a^4+2a^3b+5a^2b^2+4ab^3+6b^4),\nn
R_{MNKL}R^{MNKL}&=&\f{4}{L^4}(a^4+2a^2b^2+3b^4) \eea are that the
coordinate invariant quantities do depends on both the exponents $a$
and $b$ and $T^M_M$ is the trace of the energy-momentum tensor.

Let us do some change of coordinates\footnote{Thanks to Alex Maloney
for a useful correspondence.} and try to make contact with
\cite{klm} \be r^b=\rho, ~~(t,~x,~y)\rightarrow \f{1}{b} (t,~x,~y),
~~L\rightarrow b L.\ee This change of coordinates makes sense only
when $b \neq 0$ and under this change we get the metric
eq(\ref{metric}) as \be\label{metric_klm} ds^2=L^2[-\rho^{2\f{a}{b}}
dt^2+\rho^2(dx^2+dy^2)+\f{d\rho^2}{\rho^2}].\ee This form of the
metric coincides with the one written in \cite{klm}  by defining
$z=a/b$.

We can do a similar change of coordinates \be r^a=\widetilde{\rho},
~~(t,~x,~y)\rightarrow \f{1}{a} (t,~x,~,y), ~~L\rightarrow a L,\ee
and this makes sense when $a\neq 0$. Under this change we get the
metric eq(\ref{metric}) as \be ds^2=L^2[-\widetilde{\rho}^2
dt^2+\widetilde{\rho}^{2\f{b}{a}}(dx^2+dy^2)+\f{d\widetilde{\rho}^2}{\widetilde{\rho}^2}].\ee

Let us recall the solutions that we have in eq(\ref{solutions}),
from this it follows that in order to have  real fluxes we can only
take $b$ to vanish but not $a$. The other conditions are  $ab\geq
0$,~~ and $a\geq b$.

The metric written in eq(\ref{metric}), in the $(r,~t,~x,~y)$
coordinate system is a good one as one can consider the  $b= 0$ case
but not in the $(\rho,~t,~x,~y)$ coordinate system that is used to
write eq(\ref{metric_klm}). However, if we want to concentrate on
the situations when $b$ do not vanishes, then any of coordinate
system is good.

The physics of eq(\ref{metric})  is that we can study a combination
of both spatial and temporal scale invariance as well as temporal
scale invariance independently, whereas if we use the coordinate
system that is written in  eq(\ref{metric_klm}), then study of only
temporal scale invariance is not possible. It is important to note
that in none of coordinate system that we know of to describe only
the spatial scale invariance. It  probably makes sense to say that
in order to describe only temporal scale invariance, we need to have
two exponents in the metric rather than one and we cannot study only
the spatial scale invariance because of the reality constraint on
the fluxes.

In the next section we shall study some of the properties of the
operators in the dual field theory using the generalized form of the
gauge/gravity correspondence in which we shall keep $b$ to be
arbitrary, but while studying the 2pt correlation function of
operators that are dual to scalars, we shall use a specific choice to
$a$ and $b$.
\section{Field theory observable}

Let us consider a real scalar field $\phi$ of mass $m$ that is
propagating in the background of eq(\ref{metric}), which in the
$u=1/r$ coordinate is \be
ds^2=L^2\bigg(-\f{dt^2}{u^{2a}}+\f{dx^2+dy^2}{u^{2b}}+\f{du^2}{u^2}
\bigg) \ee

The equation of motion for a minimally coupled scalar field $\phi$
is \be\label{eom_scalar}
\p^2_u\phi-\f{a+2b-1}{u}\p_u\phi-[w^2u^{2(a-1)}+(k^2_x+k^2_y)u^{2(b-1)}+
\f{(mL)^2}{u^2}]\phi=0 \ee

In order to understand the generalized form of the  AdS/CFT
dictionary in this case, we need to find the relation between the
operator dimension $\Delta$ and the mass of the field, $m$, where,
the field $\phi$ is dual to an operator of dimension $\Delta$ for
which \bea &&\Delta(\Delta-a-2b)=m^2L^2,\nn&&
\Delta_{\pm}=b+\f{a}{2}\pm\sqrt{(b+\f{a}{2})^2+m^2 L^2}. \eea
$\Delta_+$ and $\Delta_-$ are the two roots of the first equation
with $\Delta_+ \geq \Delta_-$.

On requiring the finiteness of the Euclidean action of the scalar
field, as is done in \cite{kw} for  asymptotically AdS space time,
imposes the restriction that
 if the mass of the scalar field stays \be (m L)^2 >
1-\bigg(\f{a+2b}{2}\bigg)^2 \ee above this bound then only
$\Delta_+$ branch is allowed, whereas if the scalar field has a mass
that stays between \be -\bigg(\f{a+2b}{2}\bigg)^2 <(m L)^2 <
1-\bigg(\f{a+2b}{2}\bigg)^2 \ee then both $\Delta_-$ and $\Delta_+$
branches are allowed.

The analogue of
Breitenlohner-Freedman bound \cite{bf} for this case is
\be
(mL)^2 <-\bigg(\f{a+2b}{2}\bigg)^2
\ee
and if the mass stays below this bound then there is an instability in the
system.

It is well known that in a CFT the two point correlation of an
operator with dimension $\Delta$ goes as \be <{\cal O}(x){\cal
O}(0)> \sim \f{1}{|x|^{2\Delta}} \ee

However, for a non-relativistic CFT the two point correlator instead
of just going like a power law falloff it can get dressed by an
exponential falloff \cite{bm}. Whatever be the case, it's for sure
that there will be a power law falloff, which follows from the
translational and rotational symmetry in the spatial directions.
Upon assuming that is the case
 it just follows trivially that the two point correlation function
depends  on the parameters on which $\Delta$ depends.

In our case $\Delta$ depends on  the parameter $a$ and $b$ in the
combination $a+2b$, which means there will be two exponents in the
2-pt correlation function. For $b\neq 0$, we can rewrite the
expression to $\Delta_{\pm}$ for which $\Delta_{\pm}/b$ depends on
$a$ and $b$ in a specific way that is $a/b$ with a redefinition to
$L$. But, unfortunately, there do not looks like the presence of any
phase transitions.

According to the AdS/CFT prescription \cite{ew}, \cite{gkp}
the correlation function of operators is evaluated by differentiating the
 on shell value of the action  with respect to
 a specific boundary values of the bulk field.

For the minimally coupled scalar field with generic values of $a$
and $b$, it is not easy to compute the correlation function,
analytically. However for a specific choice of $a/b=2$, one can
solve eq(\ref{eom_scalar}). If we recall from
eq(\ref{gen_scale_inv}), the value of $a/b$ for which the action is
quadratic in fields is $a/b=2$.

For  this particular choice of $a/b$,  the normalized solution is
\bea G(u,\overrightarrow{\mathbf{k}},\omega)&=&c_1 \times
2^{\f{1}{2}[1+\f{\sqrt{4b^2+m^2L^2}}{b}]} \times e^{-\f{\omega}{2b}
u^{2b}} \times u^{2b+\f{\sqrt{4b^2+m^2L^2}}{b}} \times\nn&&
U\bigg(\f{\overrightarrow{\mathbf{k}}^2+2b\omega+2\omega\sqrt{4b^2+m^2L^2}}{4b\omega},1+\f{\sqrt{4b^2+m^2
L^2}}{b},\f{\omega}{b}u^{2b}\bigg),\nn \eea where $c_1$ is a
normalization constant and to be fixed by the condition \be
G(u\ra\epsilon,\overrightarrow{\mathbf{k}},\omega)=1 \ee and
U(a,b,z) is the confluent hypergeometric function of the second
kind.

Let us recall that the action of a scalar field
\be
S=-\f{1}{2}\int d^3 xdu \sqrt{g}(g^{MN}\p_M\phi\p_N\phi+m^2\phi^2)
\ee

which upon using equation of motion can be re-written as
\be
S=-\f{1}{2}\int d^3 x [\sqrt{g}(g^{uu}\phi\p_u\phi+m^2\phi^2)]^{\infty}_{\epsilon}
\ee

where we have introduced a regulator $\epsilon$ to regulate the
ultraviolet divergences. The last equation in momentum space can be
re-written, by introducing the sources at the boundary that is
$\phi(u,k)=G(u,k)\phi(0,k)$, where we use a condensed notation
$k_{\mu}$ to represent the three vector, $k=(w,
\overrightarrow{\mathbf{k}})$.

\be S=\int d^2 k dw \phi(0,-k){\cal F}(k) \phi(0,k) \ee and the flux
factor is \be -\f{1}{2}{\cal
F}(k)=[G(u,-k)\sqrt{g}g^{uu}\p_uG(u,k)]^{\infty}_{\epsilon} \ee Now
the two point correlation function of  operator ${\cal O}$
associated to the dual of a scalar field is calculated by
differentiating twice the action of the scalar field with respect to
the source $\phi(0,k)$ \be\label{2pt_function} <{\cal O}(-k){\cal
O}(k)>={\cal F}(k) \ee

In order to proceed further let us restrict ourselves to the mass
less sector and in this case the functions that appear in the two
point correlator are

\bea G(u,k)&=&1-\f{\omega}{2b}u^{2b}+\bigg(\alpha
\f{\omega^2}{2b^2}+\f{\omega^2}{8b^2}-\nn&&\f{\Gamma(\alpha+2)}{\Gamma(\alpha)}\f{\omega^2}{4b^2}
[-3+4\gamma+2\psi(2+\alpha)+2log(\f{w}{b})+4b~log~u] \bigg)+{\cal
O}(u)^{6b}\nn \p_uG(u,k)&=&u^{4b-1}\bigg[-\omega u^{-2b}+\f{2\alpha
\omega^2}{b}+\f{\omega^2}{2b}-\nn&&\f{\Gamma(\alpha+2)}{\Gamma(\alpha)}\f{\omega^2}{b}
[-4+4\gamma+2\psi(2+\alpha)+2log(\f{\omega}{b})+4 b~log~u] +{\cal
O}(u)^{6b-1}\bigg]\nn \sqrt{g}g^{uu}&=&L^2 u^{1-4b} \eea where \be
\alpha=-\f{1}{2}+\f{\overrightarrow{\mathbf{k}}^2}{4b\omega},\ee
$\gamma$ is the Euler-Mascheroni constant and $\psi(x)$ is the
digamma function.

 using all these ingredients into
eq(\ref{2pt_function}) we find \be <{\cal O}(-k){\cal O}(k)>={\cal
F}(k)=-\f{L^2}{2} \overrightarrow{\mathbf{k}}^2
\omega+\f{L^2}{b^3}(\overrightarrow{\mathbf{k}}^4-b^2\omega^2)[-2+log(\omega)-\psi(\f{3}{2}+\f{\overrightarrow{\mathbf{k}}^2}{4b\omega})].
\ee

Summarizing all this,  we have presented  a better coordinate system
than \cite{klm} to handle the case for which $b$ vanishes.
Generically, the constraint on the parameters $a$ and $b$ are
$ab\geq 0$ and $a\geq b$. The situation when $ab=0$, we take $b$ to
vanish but not $a$, in order to have  real fluxes. It means that
the coordinate system that we used to write eq(\ref{metric}), is
better than the coordinate system used to write eq(\ref{metric_klm}), 
in the sense that we can describe
the temporal scale invariance along with the combination of both
temporal and spatial scale invariance, independently. However, if
want to have  an action in the dual field theory to be quadratic (or
any other value) in fields means, $b\neq 0$ and it suggests that
either of the coordinate system is good.

 {\bf Acknowledgment}: It is a pleasure to thank Alex Maloney for a useful
 correspondence, also to Rabin Banerjee and S N Bose centre for basic Sciences, Kolkata
for providing a warm
 hospitality.

\end{document}